\pdfoutput=1
\documentclass[a4paper]{article}%
\usepackage{amsmath}
\usepackage{amsfonts}
\usepackage{amssymb}
\usepackage{graphicx}%
\begin{document}

\title{\textbf{Phenomenological quark-lepton mass relations and neutrino mass
estimations}}
\author{Dimitar Valev\\\textit{Stara Zagora Department, Solar-Terrestrial Influences Laboratory,}\\\textit{Bulgarian Academy of Sciences, 6000 Stara Zagora, Bulgaria}}
\maketitle

\begin{abstract}
Based on the experimental data and estimations of the charged leptons and
quarks masses, a close power law with exponent 3/4 has been found, connecting
charged leptons masses and up quarks masses. A similar mass relation has been
suggested for the masses of neutral leptons and down quarks. The latter mass
relation and the results of the solar and atmospheric neutrino experiments
have been used for prediction of neutrino masses. The obtained masses of
$\nu_{e}$, $\nu_{\mu}$ and $\nu_{\tau}$ are 0.0003 $%
\operatorname{eV}%
$, 0.003 $%
\operatorname{eV}%
$ and 0.04 $%
\operatorname{eV}%
$, respectively. These values are compatible with the recent experimental data
and support the normal hierarchy of neutrino masses.



\end{abstract}

\section{Introduction}

Decades after the experimental detection of the neutrino \cite{Reines 1953},
it was generally accepted that the neutrino mass $m_{0\nu}$ was rigorously
zero. The crucial experiments with the 50 kton neutrino detector
Super-Kamiokande found strong evidence for oscillations (and hence - mass) in
the atmospheric neutrinos \cite{Fukuda 1998}. The direct neutrino measurements
allowed to bound the neutrino mass. The upper limit for the mass of the
lightest neutrino flavor $\nu_{e}$ was obtained from experiments for
measurement of the high-energy part of the tritium $\beta$-spectrum and recent
experiments yielded $\sim2%
\operatorname{eV}%
$ upper limit \cite{Weinheimer 1999, Lobashev 1999}. As a result of the recent
experiments, the upper mass limits of $\nu_{\mu\text{ }}$and $\nu_{\tau\text{
}}$are $170$ KeV \cite{Assamagan 1996} and $18.2%
\operatorname{MeV}%
$ \cite{Barate 1998}, respectively. The Solar neutrino experiments
(\textit{SNE}) and Atmospheric neutrino experiments (\textit{ANE}) allow to
find the square mass difference $\bigtriangleup m_{12}^{2}=m_{2}^{2}-m_{1}%
^{2}$ and $\bigtriangleup m_{23}^{2}=m_{3}^{2}-m_{2}^{2}$, but not the
absolute value of neutrino masses. The astrophysical constraint of the
neutrino mass is $%
{\textstyle\sum}
m_{\nu}<2%
\operatorname{eV}%
$ \cite{Bahcall 1996}. The recent extensions of the Standard model lead to
non-zero neutrino masses, which are within the large range of $10^{-6}%
\operatorname{eV}%
\div10%
\operatorname{eV}%
$.

In the classical $SU(5)$ model the mass relations between charged leptons and
down quark masses are simply identities: $m_{e}=m_{d},$ $m_{\mu}=m_{s}$ and
$m_{\tau}=m_{b}$. The mass relations of Georgi-Jarlskog \cite{Georgi 1979}
ensue from the $SO(10)$ model and relate charged leptons and down quark
masses: $m_{e}=m_{d}/3,$ $m_{\mu}=3m_{s}$ and $m_{\tau}=m_{b}$. However, these
mass relations deviate several times, compared to experimental data. Moreover,
similar mass relations are unsuited for neutral leptons (neutrino) masses.

The seesaw mechanism naturally generates small Majorana neutrino mass $m_{\nu
}$ from reasonable Dirac mass $m_{D}$ and very heavy Majorana sterile neutrino
mass $M_{N}$, namely $m_{\nu}\sim\frac{m_{D}^{2}}{M_{N}}\ll m_{D}$. But there
are many seesaw models that differ in the scale $M_{N}$ and Dirac mass. The
Grand unified theories (\textit{GUT}) are the main candidates for seesaw
models, with $M_{N}$ at or a few orders of magnitude below \textit{GUT} scale.
Successful \textit{GUT} models should essentially generate
Cabibbo-Kobayashi-Maskawa (\textit{CKM}) quark mixing matrix \cite{Cabibbo
1963, Cobayashi 1973} and Maki-Nakagawa-Sakata (\textit{MNS}) lepton mixing
matrix \cite{Maki 1962} and predict results compatible with the data from
\textit{SNE} and \textit{ANE}. Yet, it is admitted that the predictions of the
quark-lepton mass spectrum are the least successful aspect of the unified
gauge theory \cite{Fukugita 1999, Falcone 2002}.

The purpose of this paper is to find simple and reliable quark-lepton mass
relations, based on experimental data and estimations for quark and lepton
masses. The next step is to estimate neutrino masses by means of these mass
relations and data from \textit{SNE} and \textit{ANE}.

\section{Power law approximation for the masses of charged leptons and up
quarks}

According to the Standard model, the fundamental constituents of the matter
are 6 quarks and 6 leptons. The fundamental fermions group in three
generations, having similar properties and increasing masses. The three
generations of the fundamental fermions and their masses are presented in
Table \ref{Table 1}. The estimations of quark masses are taken from
\cite{Manohar 2000} and the upper mass limits of the neutrino flavors are
taken from \cite{Weinheimer 1999, Lobashev 1999, Assamagan 1996, Barate 1998}.

%

\begin{table}[htb] \centering
\caption{Three generations of fundamental fermions and their masses
(MeV).}%
\begin{tabular}
[c]{lllllll}\hline\hline
Fermions & \multicolumn{2}{l}{$1^{st}$ generation} &
\multicolumn{2}{l}{$2^{nd}$ generation} & \multicolumn{2}{l}{$3^{rd}$
generation}\\\hline
Up quarks & $u$ & $3$ & $c$ & $1.25\times10^{3}$ & $t$ & $1.74\times10^{5}$\\
Down quarks & $d$ & $6$ & $s$ & $122$ & $b$ & $4.2\times10^{3}$\\
Charged leptons & $e$ & $0.511$ & $\mu$ & $106$ & $\tau$ & $1.78\times10^{3}%
$\\
Neutral leptons & $\nu_{e}$ & $<2\times10^{-6}$ & $\nu_{\mu}$ & $<0.17$ &
$\nu_{\tau}$ & $<18.2$\\\hline\hline
\end{tabular}
\label{Table 1}%
\end{table}%

\bigskip

A clear feature of the quark and charged lepton mass spectrum is the hierarchy
of masses belonging to different generations:

\begin{center}%
\begin{equation}
m_{u}\ll m_{c}\ll m_{t},\text{ }m_{d}\ll m_{s}\ll m_{b}\text{, }m_{e}\ll
m_{\mu}\ll m_{\tau} \label{eqn1}%
\end{equation}

\end{center}

Most likely, a similar hierarchy of masses of the neutral leptons (neutrinos)
could be anticipate $m_{\nu e}\ll m_{\nu\mu}\ll m_{\nu\tau}$. Based on the
experimental data, we search for a simple relation between the masses of
charged leptons ($m_{cl}$) and the respective up quarks ($m_{uq}$) by the
least squares. Although the linear regression $m_{cl}\approx0.0102m_{uq}$ $%
\operatorname{eV}%
$ shows close correlation, it yields electron mass many times lower than the
experimental value. After examination of other simple approximations
(logarithmic, exponential and power law) we found out that the power law fits
best experimental data:

\begin{center}%
\begin{equation}
m_{cl}=k_{0}m_{uq}^{\alpha}%
\operatorname{eV}
\label{eqn2}%
\end{equation}

\end{center}

where $k_{0}=9.33$ and $\alpha=0.749\approx3/4.$

Despite the large uncertainty of the $u-$ quark mass (from $1.5%
\operatorname{MeV}%
$ to $5$ $%
\operatorname{MeV}%
$) and $d-$ quark mass (from $3%
\operatorname{MeV}%
$ to $9%
\operatorname{MeV}%
$), the slope remains within the narrow interval from 0.683 to 0.782. Although
only three points make the approximation, the correlation coefficient is very
high ($r=0.993$) and the maximal ratio of the predicted mass in relation to
the respective experimental value is $\Delta=m_{pr}/m_{ex}=$ $1.74$ (for
muon). The predicted masses of the electron and tau lepton differ from the
respective experimental values with less than 40\%. Therefore, the mass
relation (\ref{eqn2}) could be accepted satisfactory.

\section{Mass relation for neutral leptons and down quarks and estimations of
neutrino masses}

We suggest that a mass relation similar to (\ref{eqn2}) connects the masses of
the neutral leptons ($m_{nl}$) and the respective down quarks ($m_{dq}$):

\begin{center}%
\begin{equation}
m_{nl}=km_{dq}^{\alpha}%
\operatorname{eV}
\label{eqn3}%
\end{equation}

\end{center}

where $\alpha=0.749\approx3/4$ and $k$ is an unknown constant.

For $k=k_{0}=9.33$ formula (\ref{eqn3}) yields $m_{\nu e}\approx1.13$ $%
\operatorname{MeV}%
$, $m_{\nu\mu}\approx10.84%
\operatorname{MeV}%
$ and $m_{\nu\tau}\approx153.66$ $%
\operatorname{MeV}%
$. These values are several orders of magnitude bigger than the experimental
upper limits of the neutrino masses (See Table 1), therefore $k\ll k_{0}$.
Astrophysical constraints allow to limit more closely the value of \textit{k}
since they give $m_{\nu}<\sum m_{\nu}<2$ $%
\operatorname{eV}%
$. Thus, from equation (\ref{eqn3}) we obtain :

\begin{center}%
\begin{equation}
k=\frac{m_{\nu\tau}}{m_{b}^{3/4}}<\frac{\sum m_{\nu}}{m_{b}^{3/4}}%
\sim1.21\times10^{-7} \label{eqn4}%
\end{equation}

\end{center}

\textit{ANE} \cite{Ashie 2005} determine the squared mass difference:

\begin{center}%
\begin{equation}
m_{\nu\tau}^{2}-m_{\nu\mu}^{2}\approx2.2\times10^{-3}%
\operatorname{eV}%
^{2} \label{eqn5}%
\end{equation}

\end{center}

Relation (\ref{eqn3}) yields:

\begin{center}%
\begin{equation}
\frac{m_{\nu\mu}}{m_{\nu e}}\sim(\frac{m_{s}}{m_{d}})^{3/4}\approx9.60
\label{eqn6}%
\end{equation}

\begin{equation}
\frac{m_{\nu\tau}}{m_{\nu\mu}}\sim(\frac{m_{b}}{m_{s}})^{3/4}\approx14.17
\label{eqn7}%
\end{equation}

\end{center}

\bigskip

Solving system (\ref{eqn5}) - (\ref{eqn7}) we obtain $m_{\nu e}\approx
3.4\times10^{-4}%
\operatorname{eV}%
$, $m_{\nu\mu}\approx3.3\times10^{-3}%
\operatorname{eV}%
$ and $m_{\nu\tau}\approx4.7\times10^{-2}%
\operatorname{eV}%
$. These results support the normal hierarchy of neutrino masses.

On the other hand, the Large mixing angle (\textit{LMA}) of Mikheyev - Smirnov
- Wolfenstein (\textit{MSW}) solution for \textit{SNE} \cite{Bandyopadhyay
2005} yields:

\begin{center}%
\begin{equation}
m_{\nu\mu}^{2}-m_{\nu e}^{2}\approx7.9\times10^{-5}%
\operatorname{eV}%
^{2} \label{eqn8}%
\end{equation}

\end{center}

This equation, together (\ref{eqn6}) and (\ref{eqn7}), yield $m_{\nu e}%
\approx9.3\times10^{-4}%
\operatorname{eV}%
$, $m_{\nu\mu}\approx8.9\times10^{-3}%
\operatorname{eV}%
$ and $m_{\nu\tau}\approx0.13$ $%
\operatorname{eV}%
$. These values are almost three times bigger than the values obtained by
Super Kamiokande data, therefore, they do not fit well with the latter.
However, the Small mixing angle (\textit{SMA}) \textit{MSW} solution for
\textit{SNE} \cite{Albright 2001} yields:

\begin{center}%
\begin{equation}
m_{\nu\mu}^{2}-m_{\nu e}^{2}\approx6\times10^{-6}%
\operatorname{eV}%
^{2} \label{eqn9}%
\end{equation}

\end{center}

This equation, together (\ref{eqn6}) and (\ref{eqn7}), yield $m_{\nu e}%
\approx2.6\times10^{-4}%
\operatorname{eV}%
$, $m_{\nu\mu}\approx2.5\times10^{-3}%
\operatorname{eV}%
$ and $m_{\nu e}\approx3.4\times10^{-2}%
\operatorname{eV}%
$. These values differ by less than 25\% from the results obtained above by
Super Kamiokande data, which show that according to the suggested approach,
\textit{SMA MSW} solution fits better with the \textit{ANE} than \textit{LMA
MSW}.

Thus, the obtained quark-lepton mass relations and the results of the solar
and atmospheric neutrino experiments provide to estimate the masses of
$\nu_{e}$, $\nu_{\mu}$ and $\nu_{\tau}$ of $(2.6\div3.4)\times10^{-4}%
\operatorname{eV}%
$, $(2.5\div3.3)\times10^{-3}%
\operatorname{eV}%
$ and $(3.4\div4.7)\times10^{-2}%
\operatorname{eV}%
$, respectively. These values are close to the neutrino masses ($2.1\times
10^{-4}%
\operatorname{eV}%
$, $2.5\times10^{-3}%
\operatorname{eV}%
$ and $5.0\times10^{-2}%
\operatorname{eV}%
$) found in \cite{Valev 2008} by the mass relation, connecting the masses of
four stable particles and the coupling constants of the fundamental interactions.

We could calculate constant $k$ using the most trustworthy data for neutrinos
and down quarks masses, namely $\nu_{\tau}$ and $b$-quark masses:

\begin{center}%
\begin{equation}
k=\frac{m_{\nu\tau}}{m_{b}^{3/4}}\sim2.42\times10^{-9} \label{eqn10}%
\end{equation}

\bigskip
\end{center}

%

\begin{figure}
[t]
\begin{center}
\includegraphics[
natheight=2.859900in,
natwidth=5.306500in,
height=2.904in,
width=5.3627in
]%
{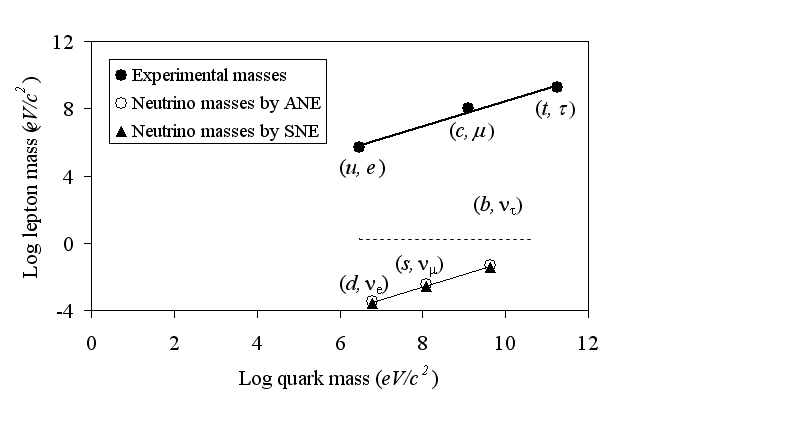}
\caption[Quark Lepton Mass]{Mass relations for the charged leptons and up
quarks masses (thick solid line) and for the neutral leptons and down quarks
masses (thin solid line). The dashed line shows the upper limit of the
neutrino masses obtained by astrophysical constraints.}%
\label{Figure1}%
\end{center}
\end{figure}

\bigskip

The obtained mass relations (\ref{eqn2}) and (\ref{eqn3}) are shown in Fig.
\ref{Figure1}. It shows that the neutrino masses estimated by \textit{SMA MSW}
are close to the neutrino masses estimated by \textit{ANE}, i.e. two sets of
estimations are compatible.

The attempt to relate the masses of the charged leptons with the masses of the
down quarks and the masses of neutral leptons with the masses of up quarks did
not yield satisfactory results since the data from \textit{SNE} and
\textit{ANE} did not fit within the framework of the suggested approach.
Besides, the respective mass relations predict the muon mass which is nearly
three times less than the experimental value (see Table \ref{Table 2}) and an
electron neutrino mass less than $10^{-7}%
\operatorname{eV}%
$.

\bigskip%
\begin{table}[htbp] \centering
\caption{Masses of charged leptons calculated by various approaches and
experimental values (MeV ).}%
\begin{tabular}
[c]{lllllll}\hline\hline
Model & \textit{SU(5)} & \textit{SO(10)} & Power law & Linear & Power law &
Exp.\\
&  &  & (Doun & (Up & (Up & data\\
&  &  & quarks) & quarks) & quarks) & \\\hline
Electron & 6 & 2 & 0.905 & 0.031 & \textbf{0.663} & \textbf{0.511}\\
Muon & 122 & 366 & 36.9 & 12.8 & \textbf{60.7} & \textbf{105.7}\\
Tau & 4200 & 4200 & 2877 & 1775 & \textbf{2449} & \textbf{1777}\\
$\Delta\max$ & 11.74 & 3.91 & 2.86 & 16.48 & \textbf{1.74} & \textbf{1}%
\\\hline\hline
\end{tabular}
\label{Table 2}%
\end{table}%

\bigskip

Table \ref{Table 2} shows the masses of charged leptons calculated by
different approaches and experimental values. The last row of the table shows
the maximal ratio (deviation) $\Delta_{\max}$ of the masses predicted by the
respective approach in relation to the experimental values $\Delta
=m_{pr}/m_{ex}$. Clearly, the power law relating to the masses of charged
leptons and up quarks fits best experimental data.

\bigskip

\section{Conclusions}

Based on the experimental data and estimations of charged leptons and quarks
masses, a power law with exponent $3/4$ has been found, connecting charged
leptons masses and up quarks masses. It has been shown that this approximation
is considerably better than any known approach. A similar mass relation has
been suggested for neutral leptons and down quarks. The latter mass relation
and the results of \textit{ANE} and \textit{SNE} have been used for
estimations of neutrino masses. The values of neutrino masses obtained by
\textit{ANE} are close to the ones, obtained by the \textit{SMA MSW} solution.
The masses of $\nu_{e}$, $\nu_{\mu}$ and $\nu_{\tau}$ are estimated to
$(2.6\div3.4)\times10^{-4}%
\operatorname{eV}%
$, $(2.5\div3.3)\times10^{-3}%
\operatorname{eV}%
$ and $(3.4\div4.7)\times10^{-2}%
\operatorname{eV}%
$, respectively, and they support the normal hierarchy of neutrino masses.


\bigskip

\begin{thebibliography}{99}                                                                                               %


\bibitem {Reines 1953}F. Reines and C. Cowan, Phys. Rev. 92 (1953) 830.

\bibitem {Fukuda 1998}Y. Fukuda et al., Phys. Rev. Lett. 81 (1998) 1562.

\bibitem {Weinheimer 1999}C. Weinheimer et al., Phys. Lett. B 460 (1999) 219.

\bibitem {Lobashev 1999}V. Lobashev et al., Phys. Lett. B 460 (1999) 227.

\bibitem {Assamagan 1996}K. Assamagan et al., Phys. Rev. D 53 (1996) 6065.

\bibitem {Barate 1998}R. Barate et al., Eur. Phys. J. C 2 (1998) 395.

\bibitem {Bahcall 1996}J. Bahcall, Astrophys. J. 467 (1996) 475.

\bibitem {Georgi 1979}H. Georgi and C. Jarlskog, Phys. Lett. B 86 (1979) 297.

\bibitem {Cabibbo 1963}N. Cabibbo, Phys. Rev. Lett. 10 (1963) 531.

\bibitem {Cobayashi 1973}M. Cobayashi and T. Maskawa, Prog. Theor. Phys. 49
(1973) 652.

\bibitem {Maki 1962}Z. Maki, M. Nakagawa and S. Sakata, Prog. Theor. Phys. 28
(1962) 870.

\bibitem {Fukugita 1999}M. Fukugita, M. Tanimoto and T. Yanagida, Phys. Rev. D
59 (1999) 113016.

\bibitem {Falcone 2002}D. Falcone, Int. J. Mod. Phys. A17 (2002) 3981.

\bibitem {Manohar 2000}A. Manohar, Eur. Phys. J. C 15 (2000) 382.

\bibitem {Ashie 2005}Y. Ashie et al., Phys. Rev. D 71 (2005) 112005.

\bibitem {Bandyopadhyay 2005}A. Bandyopadhyay et al., Phys. Lett. B 608 (2005) 115.

\bibitem {Albright 2001}C. Albright, Nucl. Instr. Meth. Phys. Res. A 472
(2001) 359.

\bibitem {Valev 2008}D. Valev, Aerospace Res. Bulg. 22 (2008) 68; http://arxiv.org/abs/hep-ph/0507255
\end{thebibliography}
\end{document}